\documentstyle[pra,aps,multicol,psfig]{revtex}

\def\addcontentsline#1#2#3{\relax}
\begin{document}
\draft
\title{Threshold features in transport through a 1D constriction. }
\author{V.V. Ponomarenko$^{1,2}$ and N. Nagaosa$^{1}$}
\address{$^1$ Department of Applied Physics, University of Tokyo,
Bunkyo-ku, Tokyo 113, Japan\\
$^2$ A.F.Ioffe Physical Technical Institute,
194021, St. Petersburg, Russia}
\date{\today}
\maketitle
\begin{abstract}

Suppression of electron current $ \Delta I$ through a 1D channel of length $L$ 
connecting two Fermi 
liquid reservoirs is studied taking into account the 
Umklapp electron-electron interaction induced by a periodic potential. 
This interaction causes  Hubbard gaps $E_H$ for $L \to \infty$.
In the perturbative regime where $E_H \ll v_c/L$ 
($v_c:$ charge velocity), and for small deviations $\delta n$
of the electron density from its commensurate values 
$- \Delta I/V$ can diverge with some exponent as 
voltage or temperature $V,T$ decreases above  $E_c=max(v_c/L,v_c \delta n)$,
while it goes to zero below $E_c$.
This results in a nonmonotonous behavior of the conductance.

\end{abstract}


%\narrowtext
%
\multicols{2}
Recent developements in the nano-fabrication technique have made the 
1D interacting electron systems an experimental reality,
and its quantum transport properties have been the subject of extensive
studies both experimentally \cite{tar,ya} and theoretically [3-15]. 
In realistic experimental set-ups, the quantum wire is attached to 
two-dimensional regions called reseivoirs or leads. 
To describe 1D transport phenomena in this configuration a model was recently
formulated of an inhomogeneous Tomonaga-Luttinger liquid
(ITTL) \cite{mas1,pon,safi}. It 
recovers the conductance $G = 2e^2/h$ observed experimentally
even in the presence of the electron-electron 
interaction in the wire \cite{tar}, although
the previous calculations on an infinitely long wire \cite{kf,of} 
predicted the renormalized conductance.
( see \cite{kawa} for the futher development). The same puzzle 
holds in the Integer Quantum Hall transport, c.f. \cite{fra}. 
Calculation of the conductance suppressed by a weak random impurity potential
in this model 
\cite{mas2} had agreed with both the previous theoretical prediction
\cite{of} and an experiment \cite{tar} (some amendment to that result
will be pointed out below).
Furthermore Tarucha et al. \cite{tar2} succeeded to introduce the 
1D periodic potential with a periodicity of 
order 40nm into the wire 2$\mu$m in length and 
50nm wide.This induces the Umklapp scatterings.
The electron density $n$ can be continuously
controlled by the gate voltage, and one can satisfy the half-filling condition 
within an accessible value of $n$.
If this condition is satisfied the system will becomes a 1D (doped) Mott
insulator with the Hubbard gap $E_H$ for the infinite length wire.
Then it will offer an idealistic system to study the 
quantum transport in Mott and doped Mott insulators in 1D.

Inspired by these works, we study  in this letter theoretically 
the $I-V$ characteristic of the wire of length $L$ connected to leads 
taking into account the effects of the    
Umklapp electron-electron interactions, 
which can be directly compared with the experiments.
We calculate the suppression of the current $ \Delta I$ 
perturbatively in the Umklapp scatterings, and the results are 
summarized in Figs. 1 and 2 for a threshold structure near half filling.
There are two energy scales, i.e., the finite size energy
$T_L = v_c/L$ ($v_c:$ charge velocity) and $E_{thr} \sim v_c \delta n$ 
($\delta n:$ the deviation of the electron density from its value at
the filling $\nu$ equal to $1/2$) 
measuring the incommensurability.
For $T,V>E_c = max(v_c/L,E_{thr})$, the suppression of the 
current $-\Delta I/V$ diverges as $max(T,V)^{4g-3}$  
if the short range
interaction constant $g$ for the forward scattering is less than $3/4$. 
For $T,V<E_c$, on the other hand, the suppression $-\Delta I/V$ goes to zero
as $T,V \rightarrow 0$.
Then we predict the nonmonotonous temperature and/or voltage dependence 
of $I$,
which is the clear signiture of the Umklapp scattering effect.
For small values of $g $, expected 
when the screening length $\xi_c$ of the interaction is much larger
than the width of the channel $d$ and $g \propto 1/\sqrt{\ln \xi_c/d} $
\cite{shul}, we predict a few more threshold singularities.
These features could be observed experimentally
by changing the gate voltage, bias voltage, and temperature. 

Our model can be derived following \cite{pon}
from a 1 channel electron Hamiltonian 
%\begin{equation}
%{\cal H}= \int d\!x \{ \sum_\sigma \psi_\sigma^+(x) ( \epsilon(
% -i \partial_x ) - E_F)\psi_\sigma(x) + \varphi(x) \rho^2(x) +
%V_{imp}(x) \rho(x) \} 
%\label{1}
%\end{equation}
\begin{eqnarray}
\lefteqn{ {\cal H}= \int d\!x \{ \sum_\sigma \psi_\sigma^+(x) ( - \frac{
\partial^2_x}{2m^*} - E_F)\psi_\sigma(x)}
\nonumber\\
& & \hspace{10mm} + \varphi(x) \rho^2(x) +
[V_{imp}(x) + V_{period}(x)] \rho(x) \} 
\label{1}
%\end{equation}
\end{eqnarray}
with the periodic potential $V_{period}(x)$
( period $a$) assumed to be weak enough to justify the perturbative
consideration of the Umklapp backscatterings.
The Fermi momentum $k_F$ and the Fermi energy $ E_F$
is determined by the filling factor $\nu$ as $\nu=k_F a/\pi$ 
and $ E_F \approx v_F k_F$.
In Eq. (\ref{1}) the function $\varphi(x)= \theta (x) \theta (L-x)$ switches on the
electron-electron interaction inside the wire confined in $0<x<L$.
Contribution of the random impurity potential $V_{imp}(x) \rho(x)$
to the conductance has been considered in \cite{of,mas2}, some results 
of which we will use below. 
 Following Haldane's 
generalized bosonization procedure \cite{hald} to account for the nonlinear 
dispersion one has to write the fermionic fields as 
$\psi_{\sigma}(x)= \sqrt{k_F/(2 \pi)} 
\sum exp\{i(n+1)(k_Fx+\phi_{\sigma}(x)/2) +i\theta_{\sigma}(x)/2 \}$
and the electron density fluctuations as $\rho(x)=\sum \rho_{\sigma}(x), \ 
\rho_{\sigma}(x)=(\partial_x \phi_{\sigma}(x))/(2 \pi) 
\sum exp\{in(k_Fx+\phi_{\sigma}(x)/2)\}$
where summation runs over even $n$ and 
$\phi_{\sigma}, \theta_{\sigma} $ are mutually conjugated bosonic fields 
$[\phi_{\sigma}(x), \theta_{\sigma}(y)]=i 2 \pi sgn(x-y)$. 

After substitution of these expressions into (\ref{1}) and introduction of 
the charge and spin bosonic fields as 
$\phi_{c,s}=(\phi_\uparrow \pm \phi_\downarrow)/ \sqrt{2}$ the Hamiltonian
takes its bose-form ${\cal H}={\cal H}_O + {\cal H}_{bs}$. Here the free 
electron movement modified by the forward scattering interaction is 
described by \cite{mas1,pon,safi}
%\begin{equation}
%{\cal H}_O= \int dx \sum_{b=c,s} \frac{v_b}{2}
%\{ { 1 \over {g_b(x)} } 
%\left({{\partial_x \phi_b(x) } \over {\sqrt{4 \pi}}} \right)^2 + 
%g_b(x) \left({{\partial_x \theta_b(x) } \over {\sqrt{4 \pi}}}
%\right)^2 \}
%\label{2}
%\end{equation}
\begin{eqnarray}
\lefteqn{{\cal H}_O= \int dx \sum_{b=c,s} \frac{v_b}{2}
\{ { 1 \over {g_b(x)} } 
\left({{\partial_x \phi_b(x) } \over {\sqrt{4 \pi}}} \right)^2 }
\nonumber \\
& & \hspace{40mm} + g_b(x) \left({{\partial_x \theta_b(x) }
     \over {\sqrt{4 \pi}}}
\right)^2 \}
\label{2}
\end{eqnarray}
%$
%{\cal H}_O= \int dx \sum_{b=c,s} \frac{v_b}{2}
%\{ { 1 \over {g_b(x)} } 
%\left({{\partial_x \phi_b(x) } \over {\sqrt{4 \pi}}} \right)^2 + 
%g_b(x) \left({{\partial_x \theta_b(x) } \over {\sqrt{4 \pi}}}
%\right)^2 \}
%\label{2}
%$
with $g_c(x)=g$ for $x \in [0,L]$ ( $g$ is less than 1 for the repulsive
interaction and it will be assumed below ), $g_c(x)=1$, otherwise and
$v_c(x)=v_F/g_c(x)$. The constants in the spin channel $g_s=1, v_s=v_F$
are fixed by $SU(2)$ symmetry.
Keeping only the most slowly decaying terms among others with the same
transfered momentum one could write the backscattering interaction as
\endmulticols
\vspace{-6mm}\noindent\underline{\hspace{87mm}}
\begin{equation}
{\cal H}_{bs}= {E_F^2 \over v_F} \int_{0}^{L} dx \bigl[
\sum_{even \  m >  0} U_m \cos(2 k_{mF} m x + 
{{m \phi_c(x)} \over \sqrt{2}}) + \sum_{odd \  m >1} U_m
\cos({\phi_s(x) \over \sqrt{2}})
cos(2 k_{mF} m x + {{m \phi_c(x)} \over \sqrt{2}}) \bigr]
\label{3}
\end{equation}
\noindent\hspace{92mm}\underline{\hspace{87mm}}\vspace{-3mm}
\multicols{2}\noindent
Difference in the transfered momentums $2 k_{mF}$ is brought up by the
periodical potential with the period $a$:
$k_{mF}=k_F -  \pi l/(m a))>0$, where $l$ is an integer chosen to minimize 
$k_{mF}$. 
We have omitted the first terms in both
sums for they could not affect the current in the lowest perturbative
order: the $m=0$ term of the first sum can contain only the spin field and 
the $m=1$ term of the second sum is assumed to  transfer large momentum. The
dimensionless coefficients $U_m$ originating from $\varphi(x)$ will be
assumed to be small enough to justify perturbative 
calculation of the current. 

Change of current due to the backscattering is given by
: $\Delta\! I=-i/(2 \sqrt{2} \pi) \int d\!x \, [\partial_x \theta_c(x), 
{\cal H}]= \sqrt{2} \int d\!x (\delta/\delta\!\phi_c(x)) {\cal H}_{bs}$. 
At finite voltage $V$ applied symmetrically to neglect the momentum
transfer variation, the average of $\Delta\! I$ 
decomposes into sum of the different backscattering mechanism contributions
$<\Delta \! I_m>$ in the lowest order.
The even $m$ terms involving only $\phi_c$ field are equal to 
%\begin{eqnarray}
%<\Delta\! I_m>= -{m \over 4} \bigl({{U_m E_F^2} \over v_F}\bigr)^2
%\int_{- \infty}^{\infty}d\!t \int\! \int_{0}^{L} d \!x_1 d \! x_2
%<e^{im \phi_c(x_1,t)/\sqrt{2}} e^{-im \phi_c(x_2,0)/\sqrt{2}}> \nonumber \\
%\bigl[e^{im(2k_{mF}(x_1-x_2)+Vt)} - h.c. \bigr].
%\label{4}
%\end{eqnarray}
%
\endmulticols
\vspace{-7mm}\noindent\underline{\hspace{87mm}}
\begin{eqnarray}
<\Delta\! I_m>= -{m \over 4} \bigl({{U_m E_F^2} \over v_F}\bigr)^2
\int_{- \infty}^{\infty}dt \int\! \int_{0}^{L} d \!x_1 d \! x_2
<e^{im \phi_c(x_1,t)/\sqrt{2}} e^{-im \phi_c(x_2,0)/\sqrt{2}}>
\bigl[e^{im(2k_{mF}(x_1-x_2)+Vt)} - h.c. \bigr].
\label{4}
\end{eqnarray}
\noindent\hspace{92mm}\underline{\hspace{87mm}}\vspace{-3mm}
\multicols{2}\noindent
The current operator $\Delta \! I_m$ has a high energy scaling
dimension $m^2 g/2$ and a free electron ($g=1$) behavior at low energy.
We will see below that the integral (\ref{4}) scales at low energy 
with $(m^2-1)$ exponent and with $(m^2 g-2)$ exponent at high energy.
The most singular behavior is
due to Umklapp backscattering at $m=2$ with the threshold voltage 
$V=E_{thr}=2k_{2F} v_c$ going to zero at the half filling. Less singular
correction with $m=4$ could become relevant at the one and three quaters fillings
 and so on.
Expressions for the odd $m$ terms include additionally a spin field
correlator $<e^{i \phi_s(x_1,t)/\sqrt{2}} e^{-i \phi_s(x_2,0)/\sqrt{2}}>$
under the integrals in (\ref{4}). 
The high energy dimension of $\Delta \! I_m$ in this case is $m^2 g/2+1/2$.
The most singular behavior occurs to
the $m=3$ term at the one and two thirds fillings. It has two threshold energies
$E_{thr\,c,s}=2 k_{3F} v_{c,s}$ for $v_c \neq v_s$.  

Correlator of the charge field exponents $e^{i \phi_c(x,t)}$,
evolution of which is specified by ${\cal H}_O$, 
could be compiled from the correlators of the uniform TL
liquid $K(x,t) = K(x,t,g,v_c)$ ($K(x,t,g,v)\equiv  
(\alpha \pi /\beta )^{2g} /( \prod_{\pm} 
\sinh^g(\pi (x/v \pm (t-i \alpha))/\beta )) )
$ in the following way \cite{mas1,mas2} 
\endmulticols
\vspace{-7mm}\noindent\underline{\hspace{87mm}}
%
%\begin{eqnarray}
%<e^{i \phi_c(x,t)} e^{-i \phi_c(y,0)}>=K(x-y,t) \prod_{\pm, m=1}^{\infty}
%\bigl( { { K(2mL,0)} \over {K(2mL \pm \mid x-y \mid ,t)} } \bigr)^{r^{2m}}
%\label{5} \\
%\prod_{ m=0}^{\infty} \bigl(
% { { K(2(mL+x),0) K(2(mL+y),0)} \over {K^2(2mL+x+y,t)} }\bigr)^{r^{2m+1}/2}
%\prod_{ m=1}^{\infty} \bigl(
% { { K(2(mL-x),0) K(2(mL-y),0)} \over {K^2(2mL-x-y,t)} }\bigr)^{r^{2m-1}/2}, 
%\nonumber \\ 
%K(x,t)\equiv K_{g,c}(x,t)= 
%\left({{(\alpha \pi /\beta )^2} \over { \prod_{\pm} 
%\sinh(\pi (x/v_c \pm (t-i \alpha))/\beta )}} \right)^g
%\nonumber
%\end{eqnarray}
\begin{eqnarray}
\lefteqn{
<e^{i \phi_c(x,t)} e^{-i \phi_c(y,0)}>=K(x-y,t) \prod_{\pm, n=1}^{\infty}
\bigl( { { K(2nL,0)} \over {K(2nL \pm \mid x-y \mid ,t)} } \bigr)^{-r^{2n}}}
\label{5} \\
& & \hspace{15mm} \times \prod_{ n=0}^{\infty} \bigl(
 { { K(2(nL+x),0) K(2(nL+y),0)} \over {K^2(2nL+x+y,t)} }\bigr)^{-r^{2n+1}/2}
\prod_{ n=1}^{\infty} \bigl(
 { { K(2(nL-x),0) K(2(nL-y),0)} \over {K^2(2nL-x-y,t)} }\bigr)^{-r^{2n-1}/2}, 
\nonumber
\end{eqnarray}
\noindent\hspace{92mm}\underline{\hspace{87mm}}\vspace{-3mm}
\multicols{2}\noindent
Here $\beta$ is inverse temperature $1/T$ and $\alpha =1/E_F$ is the 
ultraviolet cut-off. This complicated form comes about through a multiple
scattering at the points of joint $x=0,L$. As a result of the scattering the 
correlator $<\phi_c(x,t) \phi_c(y,0)>$ becomes an infinite sum of the
uniform correlators taken along the different paths connecting points
x and y and undergoing reflections from the boundaries at $x=0,L$. Each
reflection brings additional factor $r=(1-g)/(1+g)$. 
The similar correlator
$<e^{i \phi_s(x,t)} e^{-i \phi_s(y,0)}>$ for spin field is
$K(x-y,t,1,v_s)$.
Below we analyze the current
corrections (\ref{4}) for high ($T>1/t_L=T_L$) and low
($T \ll T_L$) temperatures, respectively. 

\noindent
1.{\it High temperatures} $T>T_L$ - The uniform correlator $K(x,t)$ goes down
exponentially if distance between the points $\mid x \mid$ exceeds
the inverse temperature. Therefore only paths with length less than 
$\beta $ contribute to the correlator (\ref{5}). This means that the 
high temperature form of the correlator (\ref{5}) reduces to the first
multiplier $K(x-y,t)$ up to a factor $(1+O(exp(-t_L/\beta ))$.
Neglecting $O(T_L/T)$
quantity we can extend integration over $x_1-x_2$ in (\ref{4}) from $-\infty $
to $+\infty $. Then calculation of the $m=2$  contribution
reduces to finding  Fourier transformation $F_{2g}(q,\varepsilon)$ of 
the correlator $K^2(x,t,g, v_c)$:
\endmulticols
\vspace{-7mm}\noindent\underline{\hspace{87mm}}
%
%\begin{eqnarray}
%<\Delta\! I_2>=-{1 \over 4} \left({{U_2 E_F^2} \over g}\right)^2
%t_L \sum_{\pm} \pm F_{2g}(2 E_{thr},\pm 2V)=  \label{8}\\
%-2 \bigl({2^{2(g-1)} \over {\Gamma(2g)}}\bigr)^2 {E_F^2 \over T_L}
%\bigl({{\pi T}\over E_F} \bigr)^{4g-2} \sinh({V \over T})
%\prod_{\pm,\pm} \Gamma(g \pm i{{V \pm E_{thr}}\over {2 \pi T}})
%\nonumber
%\end{eqnarray}
\begin{eqnarray}
<\Delta\! I_2>={1 \over 4} \left({{U_2 E_F^2} \over g}\right)^2
t_L \sum_{\pm} \mp F_{2g}(2 E_{thr},\pm 2V)=
-2 \bigl({{2^{2(g-1)}U_2} \over {\Gamma(2g) g}}\bigr)^2 {E_F^2 \over T_L}
\bigl({{\pi T}\over E_F} \bigr)^{4g-2} \sinh({V \over T})
\prod_{\pm,\pm} \Gamma(g \pm i{{V \pm E_{thr}}\over {2 \pi T}})
\label{8}
\end{eqnarray}
One can easily see its behavior making use of the following asymtpotics:
\begin{eqnarray}
<\Delta\! I_2> \propto -\left({U_2  \over g}\right)^2 {E_F^2 \over T_L}
\left\{ \matrix{((V^2-E_{thr}^2)/E^2_F)^{2g-1}, 
&  V \gg E_{thr},\ T \cr
((V+E_{thr}) T/E^2_F)^{2g-1},\ \  &  V \approx E_{thr} \gg T \cr } 
\right.  \label{9} \\
<\Delta\! I_2> \propto -\left({U_2  \over g}\right)^2 {E_F^2 \over T_L}
\sinh{V \over T } \left\{ 
\matrix{ e^{-E_{thr}/T}((E_{thr}^2-V^2)/E^2_F)^{2g-1}, \ \ 
& E_{thr} \gg V,\ T \cr
(T/E_F)^{4g-2}, &  V \ , E_{thr} \ll T  \cr } \right.
\nonumber
\end{eqnarray}
\noindent\hspace{92mm}\underline{\hspace{87mm}}\vspace{-3mm}
\multicols{2}\noindent
\begin{figure}[htbp]
\begin{center}
\leavevmode
\psfig{file=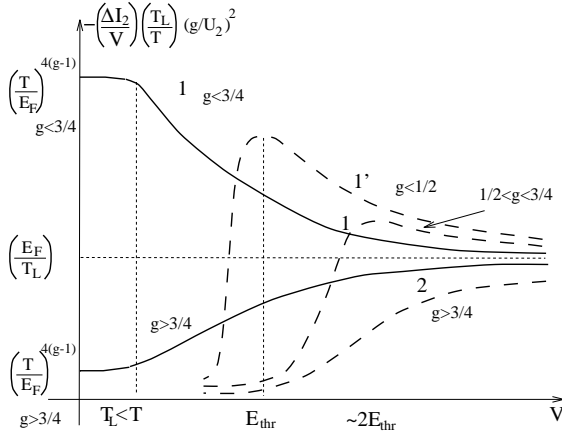,width=2.9in,angle=270}
\narrowtext{ \caption{
           Schematic voltage dependence of the high temperature
            current corrections produced by the $m=2$ Umklapp 
            interaction $\Delta I_2$. Solid lines, $E_{thr}=0$; 
            dashed lines, $E_{thr}\gg T$.  \label{cty}
           }}
\end{center}
\end{figure}
These asymptotics show that the threshold singularity in the current
voltage dependence diverges as $((V-E_{thr})/T)^{2g-1}$ if $g<1/2$
(Fig.\ref{cty}). 
It becomes stronger in the differential conductance
dependence. At $g>1/2$ the differential 
conductance correction $d\! G_2$ behaves as $-(V/E_F)^{4g-3}$, and 
saturates at $-(T/E_F)^{4g-3}$ below $T$, if $E_{thr}<T$; otherwise,
the correction shows divergence $-((V-E_{thr})/T)^{2g-2}$
smeared over $T$ scale near the threshold and 
becomes suppressed exponentially as $-exp(-E_{thr}/T)$ below it.
Generalization to the other even $m$ current corrections
needs just changing:
$4g \rightarrow m^2 g \ , 4k_{2F} \pm 2V \rightarrow m(2k_{mF} \pm V)$. 
The edge singularity is characterized
by a half of the scaling dimension for $\Delta I_m$ since only
one chiral component of the field $\phi_c$ contributes.
As to the odd $m$ terms, 
the two treshold energies
$E_{thr\,c,s}=2 k_{mF} v_{c,s}$ become distinguishable if their difference
exceeds $T$.
The leading high-temperature current correction reads as:
\endmulticols
\widetext
\vspace{-7mm}\noindent\underline{\hspace{87mm}}
\begin{equation}
<\Delta\! I_m>={ m \over 4} \left({{U_m E_F^2} \over {2 \pi g}}\right)^2
t_L v_s 
\sum_\pm \mp \int \! \int d \! q d \! \varepsilon
F_{\frac{1}{2}}(mE_{thr\,s}-q v_s, \pm mV-\varepsilon)
F_{{m^2g} \over 2}(q v_c,\varepsilon)
\label{10}
\end{equation}
\noindent\hspace{92mm}\underline{\hspace{87mm}}\vspace{-3mm}
%\widetext
\multicols{2}\noindent
Substituting zero temperature form of $F_a$ function
$F_a(q,w)=8[sin(\pi a) \Gamma(1-a)]^2 (\alpha /2)^{2a}\prod_\pm
(w\pm q)^{a-1} \theta(w\pm q)$ in (\ref{10})
one can gather that the current correction 
behaves as $-(V/E_F)^{m^2g-1}$ at large voltage $V > E_{thr\,c,s}$, has a 
leading singularity $-((V-E_{thr\,c})/T)^{m^2g/2}$ smeared over $T$ scale
near the first threshold and $-((V-E_{thr\,s})/T)^{m^2g-1/2}$ near the second
threshold (we assume $v_c>v_s$). Below the lowest threshold it becomes
exponentially suppressed. These singularities result in the divergences
of the differential conductance or higher derivatives of the current
in voltage.
The threshold behavior of the $m=3$ term of the differential
conductance correction is
divergent at $E_{thr\ c}$ if $g<2/9$ and at $E_{thr\ s}$ if $g<1/6$.

\noindent
2.{\it Low temperatures} $T,V<<T_L$ - With lowering temperature 
we should expect that above current 
correction dependences will be modulated by a $\pi T_L$ quasiperiodical
interference structure \cite{pon2,naz} and  also 
a new low energy scaling behavior 
of the current correction operators  will appear at $V,T<T_L$.
The dominant contribution to the integral of (\ref{4})
comes from long times $t \gg t_L$.  One can neglect the spacious dependence
compared with large $t$ in (\ref{5}) and keep the multipliers with the number of 
reflections $n<n^*=\beta/(2 t_L \pi )$ only to come to the long time 
asymptotics:
\endmulticols
\vspace{-7mm}\noindent\underline{\hspace{87mm}}
%
%\begin{eqnarray}
%<e^{i \phi_c(x,t)} e^{-i \phi_c(y,0)}>=e^{\gamma(T_L/T)} \bigl( {\alpha
%\over t_L } \bigr)^{2g} \bigl({{(\pi t_L/\beta )^2} \over {
%sinh(\pi (t-i\alpha)/\beta) sinh(\pi (-t+i\alpha)/\beta)}} \bigr)^{1-z}
%\nonumber \\
%\bigl({\sqrt{x y (L-x)(L-y)} \over {L^2}} \bigr)^{2rg}
%\label{6} \\
%\gamma(T_L/T)=2gr[2ln2 + \sum_{m=1}^{m^*} r^{2m-1} \ln\{4m(1+r)\}+r^{2m^*}
%\ln\{2(m^*+1)\}], \; z=r^{\beta/(t_L \pi)}
%\label{7}
%\end{eqnarray}
%
\begin{equation}
<e^{i \phi_c(x,t)} e^{-i \phi_c(y,0)}>=e^{\gamma(T_L/T)} \bigl( {\alpha
\over t_L } \bigr)^{2g} \bigl({{(\pi t_L/\beta )^2} \over {
sinh(\pi (t-i\alpha)/\beta) sinh(\pi (-t+i\alpha)/\beta)}} \bigr)^{1-z}
\bigl({\sqrt{x y (L-x)(L-y)} \over {L^2}} \bigr)^{2rg}
\label{6} 
\end{equation}
\noindent\hspace{92mm}\underline{\hspace{87mm}}\vspace{-3mm}
\multicols{2}\noindent
where $z(T_L/T)=r^{\beta/(t_L \pi)}$ and $\gamma(T_L/T)$ approach the constant
 $\gamma(\infty)$ on the order of 1 as $\ln (T_L/T) z(T_L/T)$.
Our asymptotic analysis following in essential 
Maslov's paper \cite{mas2} shows that the low energy exponents 
approach their free electron values as $ exp[T_L \ln r/(T \pi)] $.
 The effect accounts for prolongation
of the paths due to the finite reflection. 
In particular, it determines the coefficient $c(g)$ of the $T^2$
corrections to the non-universal zero temperature value of the 
conductance  variation due to impurities:
$\Delta G_{imp} \propto - (L/l) (T/E_F)^{g-1}
(1 - c(g) (T/T_L)^2)$ 
in a universal way \cite{new}. 
After substitution of (\ref{6}) into Eq. (\ref{4}),  the current suppression
produced by the even $m$ terms of the  interaction  
becomes equal to:
\endmulticols
\vspace{-7mm}\noindent\underline{\hspace{87mm}}
\begin{equation}
<\Delta\! I_m>=-{{m 2^{m^2(1-z)} e^{m^2\gamma/2 }}\over {\Gamma(m^2(1-z))}} 
\left({{U_m } \over g}\right)^2 R_{{m^2g}\over 2}(2m k_{mF}L) T_L 
\left({{\pi T } \over T_L}\right)^{m^2-1} 
\left({T_L \over E_F}\right)^{m^2(g-1)} f_{{m^2}\over 2}(V/T)
\label{11}
\end{equation}
where function 
$f_{a}(x)=\sinh(x) \prod_\pm \Gamma(a \pm ix/\pi)$
characterizes the $V-T$ cross over. 
It approaches  
$ \Gamma^2 (a)x(1 - (\ln\Gamma(a))'\!' (x/\pi)^2)$ at $ x \ll 1 $
and $\pi (x/\pi)^{2a-1}$ at $x \gg 1 $.
Function $R$ specifies the $k_F - 1/L$ cross over as:
\begin{eqnarray}
R_{2g}(x)= {{\pi \Gamma^2(1+2rg)} \over x^{1+4rg}} J^2_{1/2+2rg}(x/2) 
\simeq \Gamma^2(1+2rg)
\left\{ \matrix{ \pi/(4^{1+4rg} \Gamma^2(3/2+2rg)), \ \ 
&  x \ll 1 \cr
4 sin^2(x/2-\pi r g)x^{-2-4rg},\ \  & x \gg 1  \cr } 
\right. 
\label{12} 
\end{eqnarray}
\noindent\hspace{92mm}\underline{\hspace{87mm}}\vspace{-3mm}
\multicols{2}\noindent
\begin{figure}[htbp]
\begin{center}
\leavevmode
\psfig{file=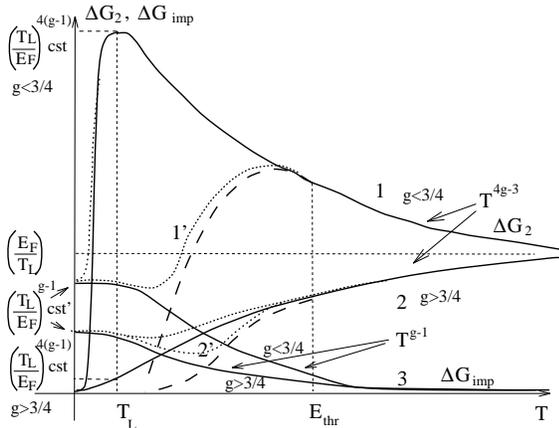,width=2.9in,angle=270}
\narrowtext{
  \caption{Schematic temperature dependence of the conductance 
         corrections produced by the $m=2$ Umklapp interaction 
         $\Delta G_2$ at $g<3/4$ (lines1)
         and at $g>3/4$ (lines2) and by the random impurity 
         potential $\Delta G_{imp}$ \cite{of,mas2} 
         (lines 3).  
         Solid lines, $E_{thr}=0$; dashed lines, $E_{thr}\gg T_L$.
         The dot lines 1 and 2 (1' and 2') are the full conductance 
         correction at $E_{thr}=0$ (finite $E_{thr})$.   \label{cn}
        }}
\end{center}
\end{figure}

The odd $m$ corrections will meet Eq.(\ref{11}) after substitution
$m^2+1$ instead of $m^2$. Combining 
above results we can outline a temperature 
dependence of the conductance 
correction produced by the $m=2$ Umklapp interaction (Fig.\ref{cn}).
Its magnitude increases/decreases
following $(E_F/T_L)(T/E_F)^{4g-3}$ as $T$ going down above $T_L$ 
and follows $(T/T_L)^{2} (T_L/E_F)^{4g-4}$, if  
$E_{thr}<T_L$; otherwise,
the correction starts to decrease exponentially  $exp(-E_{thr}/T)$
below $E_{thr}$ and keeps on decreasing like 
$(T/T_L)^2 (T_L/E_F)^{4g-4} (T_L/E_{thr})^{2+4rg} 
\sin^2(2k_{2F} L-\pi rg)$
 below $T_L$. The $T(>T_L)$ dependence is similar to that of the 
{\it conductivity} of infinite wire found by Giamarchi \cite{g}.
 Similar dependence with $T$ replaced by $V$ could be
predicted for the zero temperature differential conductance $d G_2(V)$ 
at $V<T_L$. 

In summary under perturbative condition
we have described a hierarchy of the threshold features
produced by the Umklapp backscatterings at
the rational values of the occupation number inside the 1D channel
connecting two Fermi liquid reservoirs. In the differential
conductance (its derivative) vs. volatage, the threshold 
structure is an asymmetric peak of width $ max(T,T_L)$ 
located at treshold, $E_{thr}$. In the conductance vs. temperature,
we predicted a maximum below $E_{thr}$ due to crossover from the 
Umklapp backscattering to the impurity suppression and an asymmetric 
minimum at $E_{thr}$ if the interaction is strong enough. 

The authors acknowledge H. Fukuyama, S. Tarucha for useful discussions.
This work was supported by the Center of Excellence  and 
partially by the fund for the development of collaboration
between the former Soviet Union and Japan at the JSPS.

\end{document}